\documentclass[11pt,twoside]{article}
\usepackage{aussois2003,epsf}

\def\emphasize#1{{\sl#1\/}}
\def\arg#1{{\it#1\/}}
\let\prog=\arg
\def\edcomment#1{\iffalse\marginpar{\raggedright\sl#1\/}\else\relax\fi}
\reversemarginpar
\begin{document}
%
\title{The Basics of Lensing}
%
\author{Konrad Kuijken}
\affil{Leiden Observatory, PO Box 9513, 2300RC Leiden, The Netherlands}

\label{page:first}
\begin{abstract}
The basic equations and geometry of gravitational lensing are
described, as well as the most important contexts in which it is
observed in astronomy: strong lensing, weak lensing and microlensing.
\end{abstract}
\section{Introduction}
Gravitational lensing, the subject of this volume, is the name given
to the phenomena that result from the bending of light rays by
gravitational fields. The term `lensing' is actually a little
misleading: few opticians would be satisfied with the quality of
lensing that gravitational lenses provide! Nevertheless, as the
contributions to this winter school richly illustrate, gravitational
lenses have grown into a very useful and powerful tool in astronomy
over the last decade or two.

The aim of this lecture is to lay the foundations for what follows. It
will highlight the essential aspects of light bending by gravitational
fields, and illustrate the effect this has in the most
commonly studied situations: strong lensing, weak lensing, and
microlensing. However, a review of the results obtained with
gravitational lensing is left to the other lecturers.

Much has already been written on gravitational lensing. An excellent
in-depth description of the subject is given in Schneider, Ehlers and
Falco (1992).

\section{Basic equations}

\subsection{Deflection of light by a point mass}

\begin{figure}
\plotone{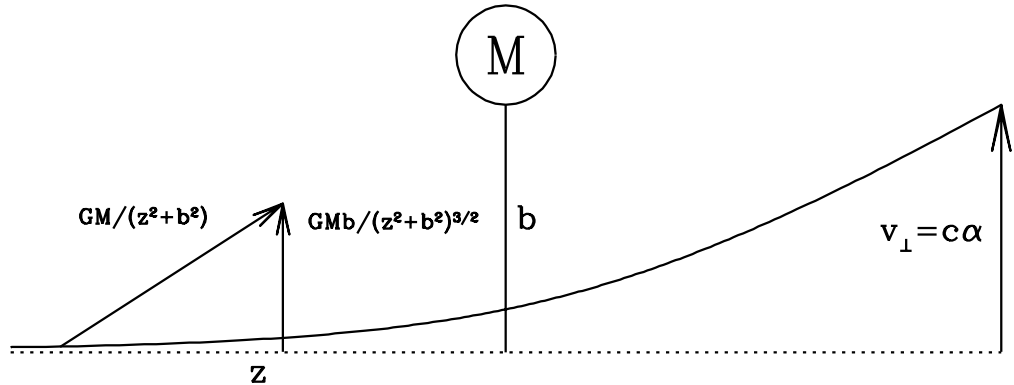}
\caption{Deflection of a light ray passing within a distance $b$ of a
  point mass $M$.}
\label{fig:ptmass}
\end{figure}

The starting point is to consider a light ray which passes close to a
point mass (Fig.~\ref{fig:ptmass}). A ray that passes within a
distance $b$ (the impact parameter) of this mass feels a Newtonian acceleration
component perpendicular to its direction of motion of 
\begin{equation}
g_\perp={GMb\over\left(b^2+z^2\right)^{3/2}}
\end{equation}
(provided the deflection is small) which results in a total integrated
velocity component $v_\perp=\int g_\perp dt = \int g_\perp dz/c = 2
GM/bc$. The resulting deflection angle is then 
\begin{equation}
\alpha = v_\perp/c = {2 GM \over bc^2} \qquad\qquad\hbox{(Newton)}.
\end{equation}
Now, it turns out that General Relativity (GR) predicts exactly twice
the deflection angle of Newtonian theory---it was in fact this factor
of two that was used as a test of GR during Eddington's solar eclipse
expedition of 1930---so that the deflection angle for a ray with
impact parameter $b$ near a point mass $M$ is
\begin{equation}
\alpha = {4 GM \over bc^2} \qquad\qquad\hbox{(GR)}\qquad\qquad\hbox{provided $\alpha\ll 1$}.
\end{equation}

In typical astronomical situations, this angle is indeed very small;
only near compact objects (white dwarfs, neutron stars, black holes)
does the deflection angle reach values of well over a minute of arc
(Table~\ref{tab:angles}). 

\begin{table}
\caption{Typical deflection angles for rays of light passing close to
  the surface of a sun-like star, to the Galaxy, or near a Galaxy cluster.}
\label{tab:angles}
\centerline{
\begin{tabular}{lccc}
\tableline
 & mass ($M_\odot$) & size (pc) & $\alpha$ (arcsec)\\
\tableline
Sun & 1 & $10^{-7}$ & 1 \\
Galaxy & $10^{11}$ & $10^4$ & 1 \\
Galaxy Cluster & $10^{14}$ & $10^5$  & 100 \\
\tableline
\tableline
\end{tabular}
}
\end{table}

\subsection{Extended lenses}

The same description can be repeated for extended mass distributions
$\rho(x,y,z)$, provided that it is still true that all lateral
acceleration takes place before the light ray has appreciably deviated
from its path (the `thin lens' approximation). In that case we can
derive the accelerations from the Newtonian potential $\psi$ of the
lens's mass distribution: assuming the light ray propagates along the
$z$ axis, we can write the lateral velocity acquired due to the lens
as
\begin{equation}
\left(\!\!\begin{array}{cc}v_x\\v_y\end{array}\!\!\right) =
2\int \left(\!\!\begin{array}{cc}\partial/\partial x\\
  \partial/\partial y\end{array}\!\!\right) \psi\, dz/c 
\end{equation} 
(where we have allowed for the factor of 2 Newton $\to$ GR). If we now
define the {\em projected potential} $\Psi(x,y)=\int\psi\,dz$, we
obtain the two-dimensional deflection angle
$(\alpha_x,\alpha_y)=(v_x,v_y)/c$ as
\begin{equation}
\hbox{\boldmath$\alpha$} ={2\over c^2}\nabla\Psi(x,y).
\label{eq:defl}
\end{equation}
Note that $\Psi$ satisfies the two-dimensional Poisson equation
\begin{equation}
\nabla^2\Psi(x,y)=4\pi G \Sigma(x,y)
\qquad\qquad\hbox{where}\qquad
\Sigma(x,y)=\int\rho(x,y,z)\,dz
\end{equation}
is the mass surface density in the lens. This equation gives the
recipe for calculating the deflection angle experienced by a light ray
that passes through the lens plane at position $(x,y)$. 

\subsection{The ray-trace equation}

\begin{figure}
\plotone{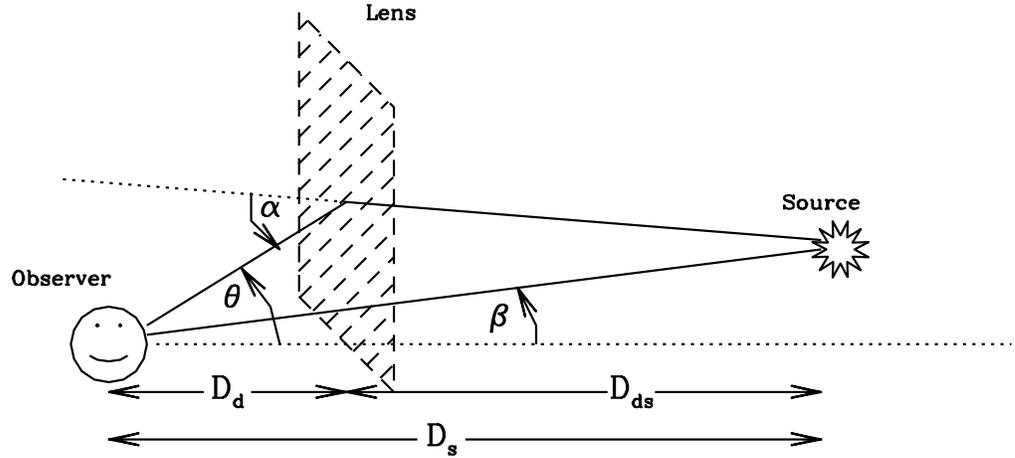}
\caption{The geometry of gravitational lensing. A source is located a
  distance $D_s$ from the observer, an angle $\beta$ away from a
  reference point in the 
  lens (deflector) which lies at distance $D_d$. Because of the
  deflection $\alpha$, the light from the source appears to come from
  direction $\theta$.}
\label{fig:raytrace}
\end{figure}

The deflection angle of equation~\ref{eq:defl} now needs to be related
to the geometry of the location of source, lens and observer, to
determine the imaging properties of a gravitational lens. The relevant
angles are shown in Fig.~\ref{fig:raytrace}.

Application of the sine rule in the triangle
observer-source-deflection point, combined with the 
approximation $\alpha, \theta, \beta\ll1$, yields
$\alpha/D_s = (\theta-\beta)/D_{ds}$, or
\begin{equation}
\beta=\theta - {D_{ds}\over D_s}\alpha(D_d\theta),
\label{eq:raytrace}
\end{equation}
which is known as the ray-trace equation. It allows a light ray to be
traced back from direction $\theta$ at the observer, past the
deflection at the location $D_d\theta$ in the lens plane, to the
source plane, so that the image of the lensed sky can be built up
pixel by pixel (surface brightness of the source is preserved under
gravitational lensing). Note that the reverse, to find the image
direction $\theta$ corresponding to a given source direction $\beta$,
is a much more difficult problem because of the generally complicated
dependence of the deflection angle on $\theta$, and may even have
multiple solutions.

As a simple example, consider a Plummer model (a softened point mass) lens. 
This model has the potential and density
\begin{equation}
\psi(r)=-{GM\over\left(r^2+a^2\right)^{1/2}}; \qquad
\rho(r)={3Ma^2\over4\pi\left(r^2+a^2\right)^{5/2}}; \qquad
\Sigma(s)={M a^2\over\pi\left(a^2+s^2\right)^2}.
\end{equation}
$M$ is the total mass of the model, and the parameter $a$ represents
the core radius of the mass distribution.
Using the results from the previous section we deduce a deflection
angle, 
as a function of the projected radius
$s=(r^2-z^2)^{1/2}$, of 
\begin{equation}
\alpha(s)={2\over c^2}\int_{-\infty}^\infty {\partial\psi\over\partial
  s}\,dz
 = {2\over c^2}\int {G M s \over \left(a^2+s^2+z^2\right)^{3/2}}\,dz
 = {4GM\over c^2}{s\over a^2+s^2}.
\end{equation}

The lens equation thus becomes (using $s=D_d \theta$)
\begin{equation}
\beta=\theta-{D_{ds}\over D_s}\alpha=
\theta-{4 GM\over c^2} {D_{ds}\over D_d D_s} {\theta\over
  \left(a/D_d\right)^2+\theta^2}.
\label{eq:lensplummer}
\end{equation}
For a given source position $\beta$, this is a cubic equation for the
image position $\theta$; it can thus have one or three roots. For a
source exactly behind the lens ($\beta=0$) the equation splits into
two factors: 
\begin{equation}
\theta=0 \qquad \hbox{or} 
\qquad \theta^2+\left(a/D_d\right)^2={4 GM\over c^2} {D_{ds}\over D_d D_s}
\end{equation}
with the second equation having real roots provided
\begin{equation}
{M\over\pi a^2}\equiv\Sigma(s=0) > {c^2\over4\pi G} {D_s\over D_d
  D_{ds}}\equiv\Sigma_{\rm crit}.
\label{eq:sigcrit}
\end{equation}
The {\em critical density} $\Sigma_{\rm crit}$ will return later; any
lens with a surface mass density above this value can 
generate multiple images. Note that the critical density decreases
when the source distance is increased: it is easier to generate
multiple images of more distant sources because the required bending
angles get smaller as the source is placed further away. At a given
source distance, the strongest lensing effect (lowest $\Sigma_{\rm
  crit}$)  is obtained with the lens roughly halfway between observer
and source.

By symmetry, the image of a source exactly behind an axisymmetric lens
will be ring-shaped---this is known as an {\em Einstein Ring}.

\begin{figure}
\plotone{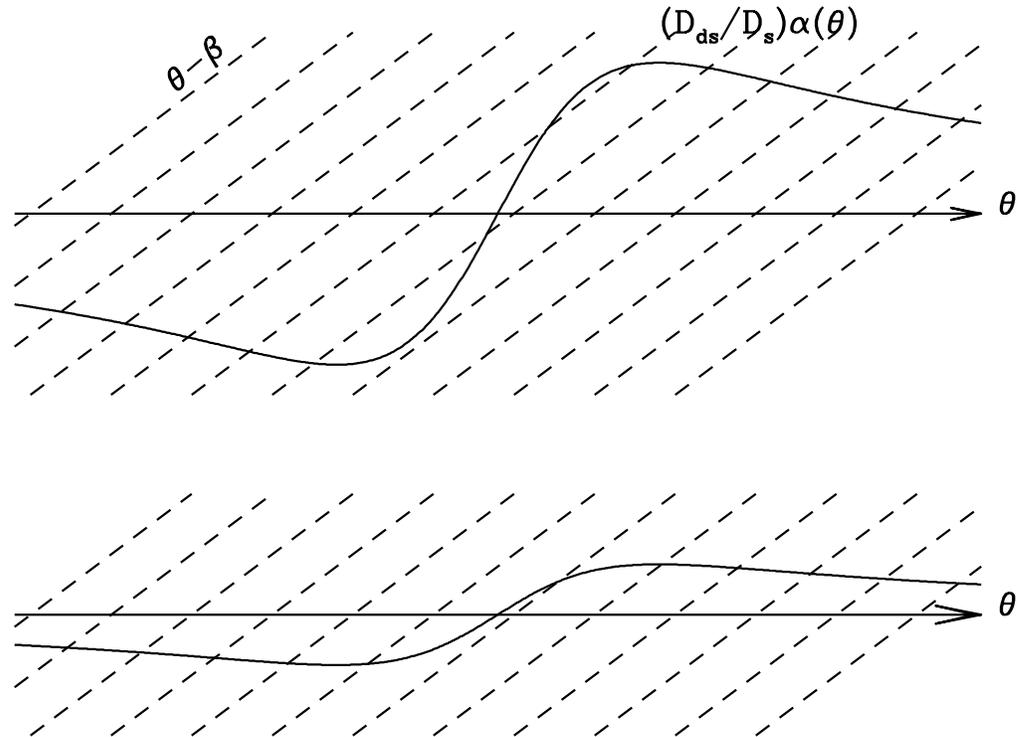}
\caption{Graphical solution of the lens equation for a Plummer model
lens. The solid line shows the deflection angle term
$(D_{ds}/D_s)\alpha$ as a function of projected angle from the lens
$\theta$. The diagonal dashed lines show the term $\theta-\beta$ for
various source locations $\beta$. The intersection of the two sets of
curves give the solutions $\theta$ to the lens equation, i.e. the
locations of the images of the source as seen through the lens. At the
top, a high central mass density lens is shown, which generates
multiple images of sources sufficiently close to the line of sight to
the lens; below the situation for a lighter lens is plotted,
and here it is clear that all sources are imaged only once.}
\label{fig:lensplummer}
\end{figure}

Returning to the lens equation for the Plummer model
(eq.~\ref{eq:lensplummer}), the structure of the solution(s) can be
illustrated graphically (Fig.~\ref{fig:lensplummer}). For a
sufficiently high central lens surface mass density (equivalent to the
central slope $d\alpha/d\theta$) multiple images can be formed of sources
that project close to the lens; otherwise only a single image is
formed, which is always further away from the lens than it would have
been in the absence of lensing (since $\alpha$ is positive).

\subsection{Arrival Time Delay and Fermat Potential}

The geometry of lensing by more complicated mass distributions is most
elegantly described by means of the {\em Fermat potential} formalism.
Fermat's principle states that light rays follow paths which represent
stationary points in arrival time---in terms of wave optics, this
corresponds to coherent phases along nearby paths, resulting in a
positive interference. 

\begin{figure}
\plotone{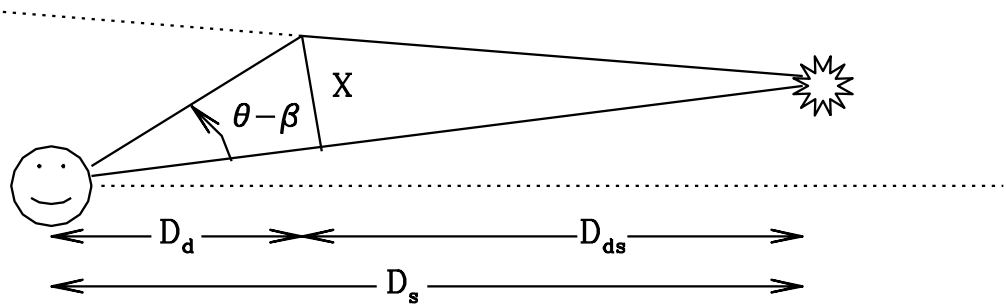}
\caption{The geometry of the arrival time calculation. The distance
  $X$ is used as an intermediate step in the calculation of the path
  length difference, using the fact that the hypotenuse of a
  rightangled triangle with base $D$ and height $X$ is equal to
  $D+X^2/(2D)$ when $X\ll D$.}
\label{fig:timedelay}
\end{figure}

In a weak gravitational field ($\psi\ll c^2$), the space-time metric
can be written as
\begin{equation}
ds^2\simeq\left(1+2\psi/c^2\right)c^2 dt^2 -
\left(1-2\psi/c^2\right)d{\bf x}^2.
\end{equation}
For a light ray $ds=0$, and hence such a ray propagating along the
$z$-axis will satisfy
\begin{equation}
c\,dt\simeq\left(1-2\psi/c^2\right)\,dz
\end{equation}
The arrival time $\int (dt/dz) dz$ for this light ray thus contains two terms: 
the path length $z/c$, and the {\em Shapiro delay} 
$-{2\over c^3}\int\psi\,dz$. If we add up both terms along a path that
has been `broken' as a result of a deflection by a gravitational lens,
we obtain (see Fig.~\ref{fig:timedelay}) a path length difference with
respect to the unlensed path of
\begin{equation}
\delta l \simeq {X^2\over2D_d} + {X^2\over2D_{ds}}
={D_s\over2D_dD_{ds}}X^2
={D_dD_s\over2 D_{ds}}\left(\beta-\theta\right)^2
\end{equation}
and a Shapiro delay of
\begin{equation}
-{2\over c^3}\int \psi\, dz = -{2\over c^3}\Psi(D_d\theta)
\end{equation}
so that the combined {\em time delay} with respect to the absence of
the lens, for a path that emanates from a source in direction $\beta$
via a lens plane position in direction $\theta$ is given by
\begin{equation}
c\delta t={D_dD_s\over 2D_{ds}}\left(\beta-\theta\right)^2 
-{2\over c^2}\Psi(D_d\theta)
\equiv \Phi(\beta, \theta)
\end{equation}
The quantity $\Phi$ is knows as the {\em Fermat potential}. Fermat's
principle states that images are seen at the locations where $\Phi$ is
stationary with respect to varying paths, i.e., where
$\nabla_\theta\Phi=0$. It is a simple exercise to show that this
corresponds to the ray-trace equation~\ref{eq:raytrace}.

Extended sources are distorted by the lens mapping: light rays
emanating from nearby points in the source plane will be deflected
slightly differently. This distortion is measured by studying the
Jacobian, or {\em magnification matrix} 
\begin{equation}
A_{ij}\equiv{\partial\theta_i\over\partial\beta_j} =
\left(\delta_{ij}-{2\over c^2}{D_{ds}D_d\over
  D_s}\Psi_{ij}\right)^{-1}
\quad\propto
\left(\partial^2\Phi\over\partial\theta_i\partial\theta_j\right)^{-1}
\end{equation}
where the $\Psi_{ij}$ are  the spatial second derivatives of the
projected potential, and $\delta_{ij}$ the Kronecker delta.
The matrix $A$ is usually written as
\begin{equation}
A = \left(\!\!\begin{array}{cc}1-\kappa-\gamma_1 & -\gamma_2\\
                               -\gamma_2& 1-\kappa+\gamma_1
               \end{array}
         \!\!\right)^{-1}
\quad\hbox{where}\quad
 \left(\!\!\begin{array}{c}\kappa\\\gamma_1\\\gamma_2\end{array}\!\!\right)
=
 {D_{ds}D_d\over c^2 D_s }
 \left(\!\!
  \begin{array}{cc}
    \Psi_{11}+\Psi_{22}\\\Psi_{11}-\Psi_{22}\\2\Psi_{12}
  \end{array}
 \!\!\right).
\label{eq:kapgam}
\end{equation}
Of course $A$ depends on the location in the image plane.  $\kappa$ is
known as the {\em convergence} of the mapping.  Note that it is given
by the divergence of the projected potential and hence proportional to
the surface mass density of the lens: in fact
$\kappa=\Sigma/\Sigma_{\rm crit}$ as defined in eq.~\ref{eq:sigcrit}.
$\gamma_1$ and $\gamma_2$ affect different directions in the image
plane differently, and are called the {\em shear} of the lens mapping.

The magnification $\cal M$ of the lens mapping is given by the determinant of
the Jacobian:
\begin{equation}
{\cal M}=\hbox{det}\left(A\right) = 
\left( (1-\kappa)^2-\gamma_1^2-\gamma_2^2 \right)^{-1}.
\label{eq:magn}
\end{equation}
If the magnification is everywhere finite, the mapping from source to
image plane is invertible and there is only one image of every
source. However, if there are multiple images of any part of the
source plane, then the mapping is no longer invertible, and the
determinant of $A^{-1}$ needs to pass through zero at some
point---this corresponds to infinite magnification. From
eq.~\ref{eq:magn} it is easy to see that a sufficient (though not
necessary) condition for infinite magnification, and hence for
multiple imaging, is for $\kappa$ to attain the value 1 somewhere
(since at large angles $\kappa$ and $\gamma_i$ tend to zero). This
corresponds to $\Sigma=\Sigma_{\rm crit}$.

In case the lens is weak ($\kappa$ and $\gamma_i\ll1$), the
magnification is approximately given by $1+2\kappa$, which is always
positive. This is a consequence of the focusing effect of the mass
over-density in the lens.

All the information of a lens mapping is thus contained in the Fermat
potential. For a given $\beta$, the stationary points of
$\Phi(\beta,\theta)$ delineate the images of a source in the direction
$\beta$. The second derivatives of $\Phi$ form the inverse
magnification matrix $A^{-1}$. At minima and maxima of $\Phi$ the
parity of the source is preserved; at a saddle point we see the source
inverted. 

One can visualize the action of a lens by considering how the Fermat
potential changes as the lens is slowly `switched on', e.g. by
changing its mass or modifying the source distance. The shape of
$\Phi$ will then gradually change from the initial, no-lensing
parabolic shape to a more complicated one: every lensing mass will
generate a positive peak in this parabola. Initially the unlensed
image will be slightly perturbed and `pushed' away, downhill, from the
lens, and slightly magnified. As the lens strength is increased,
further stationary points may be formed in the Fermat potential; these
always appear as a combination of a saddle point and a maximum or
minimum, as an extra `ridge' is created in the surface. Where a ridge
is formed, the curvature of the surface passes through zero and the
magnification of a point source is momentarily infinite before the two
new images, of opposite parity, are created and move away from the
ridge. New image pairs are therefore generated (and destroyed) near
the critical lines.

This very geometric description of the image morphology allows a
number of general laws of lensing to be stated. They apply to the case
of smooth lensing potentials (and not, for example, to a point mass
lens).

\begin{enumerate}
\item There is always at least one minimum of $\Phi$, and hence at
  least one image.
\item The total number of finite-magnification images is always odd.
\item The number of even-parity images is always one more than the
  number of odd-parity images, and new images are formed in pairs of
  opposite parity.
\end{enumerate}

\subsection{Distances}

\begin{figure}
\plotone{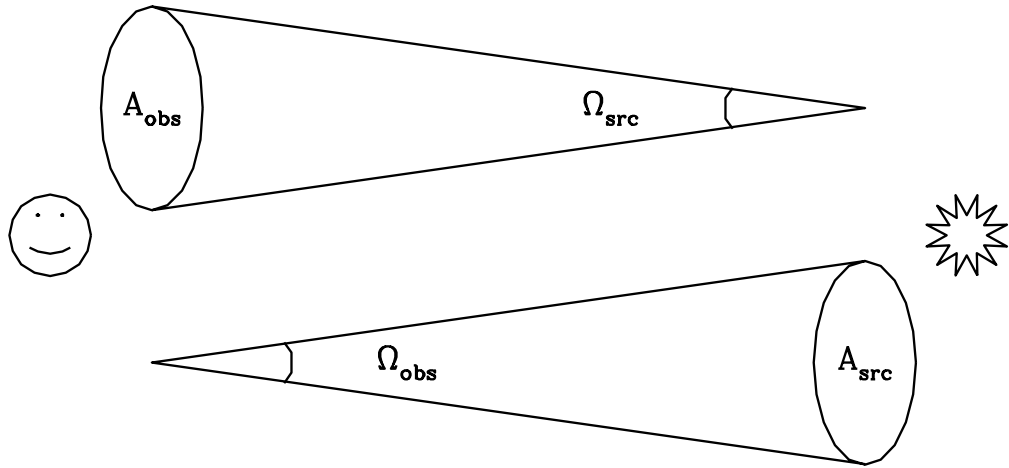}
\caption{Definitions of angular diameter distance and corrected
  luminosity distance in cosmology, as the square root of the ratio
  between appropriate area elements and the associated solid
  angle. The cone sketched at the top is used to define corrected
  luminosity distance $(A_{\rm obs}/\Omega_{\rm src})^{1/2}$, while
  the lower one is used for angular diameter distance $(A_{\rm
  src}/\Omega_{\rm obs})^{1/2}$.  }
\label{fig:angdiamdist}
\end{figure}

Thus far all derivations have been in a flat background metric, and
this makes the definition of a distance straightforward. However, in
dynamic cosmologies such as our universe, one needs to be
careful. Distance may be defined generically as $(A/\Omega)^{1/2}$
where $A$ is the size of an area element perpendicular to the line of
sight, and $\Omega$ the solid angle it subtends, but it is possible to
make different choices for $A$ and $\Omega$ (see
fig.~\ref{fig:angdiamdist}).  All results shown above remain valid in
dynamic cosmologies provided the {\em angular diameter distance} is
used. 

\subsection{Caustics and Critical Curves}

It is straightforward to apply the lens mapping: it is simply a
question of defining the surface brightness in the source plane, and
ray tracing using eq.~\ref{eq:raytrace} through the image plane to this
source plane. The mapping is only as complicated as the gradient of
the projected lens potential $\Psi$.

\begin{figure}
\plotfiddle{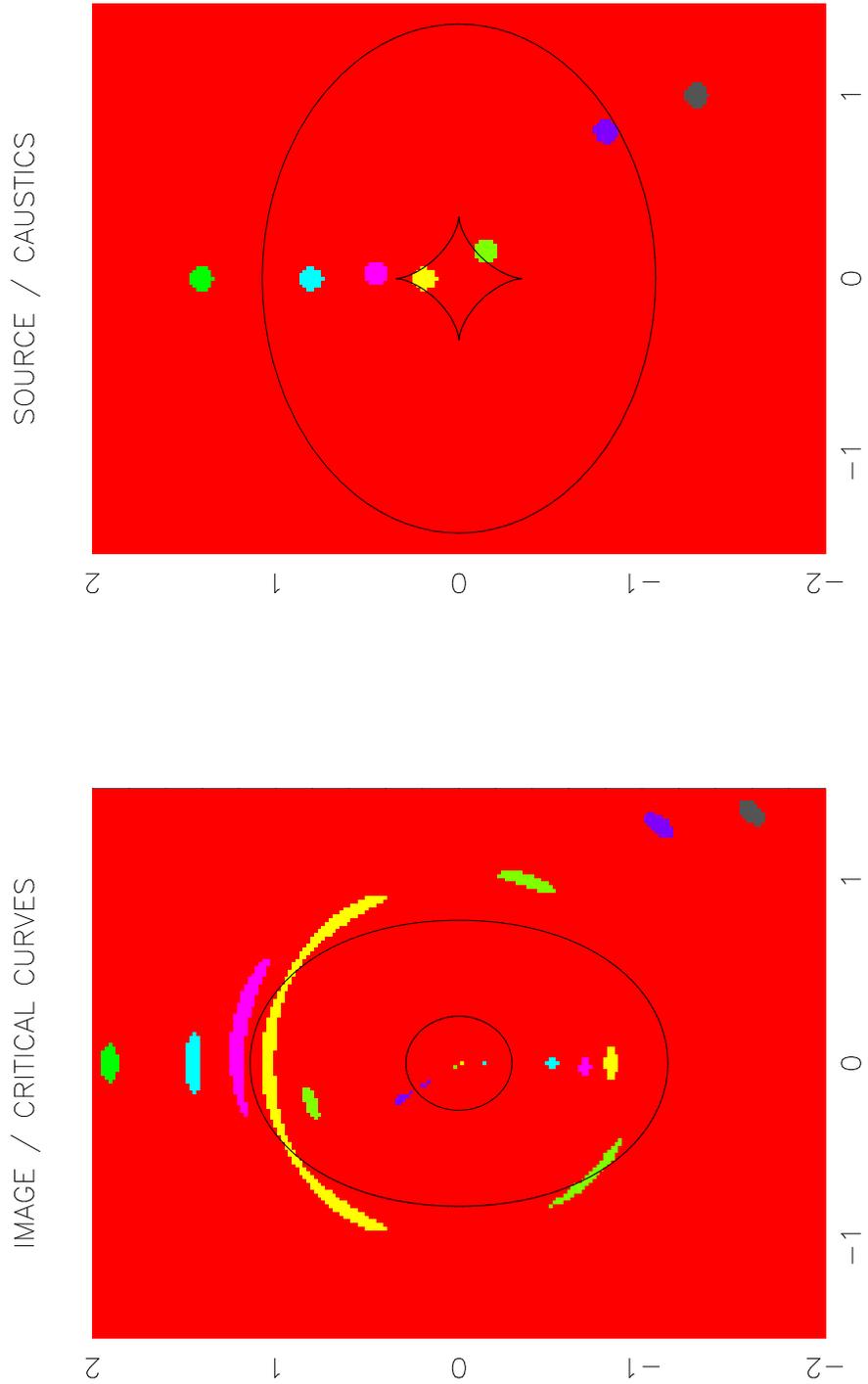}{0.85\vsize}{0}{80}{80}{-240}{-40}
\caption{Imaging of the source plane on the right through a simple
  lensing potential of the form $\Psi=\frac12\ln(a^2+x^2+Qy^2)$,
  corresponding to a flattened mass distribution with an inverse-cube
  radial density profile. }
\label{fig:lensmap}
\end{figure}

Fig.~\ref{fig:lensmap} shows an example of such a ray tracing
calculation, using a simple lens potential. On the right a number of
round sources have been defined; on the left this source plane is seen
as mapped through the lens. A number of features are clear. The lens
does two things: it pushes the images outward, and generates new
images in the interior. The outer images are tangentially distorted
(radially squeezed) by the lens.

In the left-hand panel, the {\em critical curves} of infinite
magnification ($\det(A^{-1})=0$) have been drawn. It is clear that
these mark the regions where new image pairs are generated. These
curves are as smooth as the lensing potential, and are the place to
look if one wants to see highly magnified sources.

In the right-hand panel the critical curves have been mapped back to
the source plane, where they form the {\em caustics}. They are the
locations in the source plane where multiple light rays traced back
from the observer bunch up. Because of the way these are constructed,
these need not be smooth, though they may be. Any source which falls
near a caustic is highly magnified, and when a source crosses a
caustic a pair of images is created or destroyed. The area of the
caustics is important for evaluating lensing statistics.

The inner, diamond-shaped caustic corresponds to the outer critical
line. Sources that fall near this caustic are tangentially stretched,
and if they cross it they will spawn new images near the outer
critical line. This diamond-shaped cusp is generic to elliptical
lenses, and its area increases with the flattening of the lens.

The outer caustic maps to the inner critical curve, which is where
radial arcs form. Its existence implies a core to the mass
distribution as it marks the region where the radial component of the
deflection angle peaks. Very concentrated lenses do not form radial
arcs.

\subsection{The mass-sheet degeneracy}

Even though we can learn a lot from identifying multiple image pairs
seen through a gravitational lens, it turns out to be impossible {\em
in principle} to deduce the lensing potential completely from such
data, no matter how detailed. The reason is that we do not have the
luxury of being able to remove the lens and see the source plane
undistorted. This unknown source plane propagates in the form of
degeneracies in the lens model.

The most fundamental important degeneracy stems from the unknown scale
of the source plane. If in the ray-trace equation~\ref{eq:raytrace} we
scale the source location by a factor $k$, it is simple to construct a
new lens which preserves all the image locations, as follows:
\begin{equation}
\beta=\theta-{D_{ds}\over D_s}\alpha(\theta)
\quad\Rightarrow\quad
k\beta=\theta-{D_{ds}\over D_s}
\left[k\alpha(\theta)+{D_s\over D_{ds}}(1-k)\theta\right]
\equiv \theta-{D_{ds}\over D_s}\overline\alpha(\theta).
\end{equation}
Any source at location $\beta$ viewed through the lens with deflection
angles $\alpha$ will be seen at the same location $\theta$ as a source
at location $k\beta$ seen through the lens $\overline\alpha$. This new
lens is obtained by rescaling the amplitude of the original lensing
potential by a factor $k$, and adding a new quadratic lens potential
\begin{equation}
\Psi_{\rm sheet}={c^2\over4}(1-k){D_s\over D_d D_{ds}}s^2,
\end{equation}
which corresponds to a sheet
of constant surface mass density $\Sigma=(1-k)\Sigma_{\rm crit}$. 

This {\em mass sheet degeneracy} can only be broken if (i) there is
independent knowledge of the scale of the source plane, e.g., via
source densities, (ii) sources at very different distances behind
the lens are multiply imaged, so that the distance dependence of
$\Sigma_{\rm crit}$ can be used to solve for $k$, or (iii) the lens
model can be extended out to sufficiently large radii that a mass
sheet can be ruled out on other grounds.

\section{Strong lensing}

Lenses in which $\kappa$ and $\gamma$ are of order 1 are termed {\em
strong}. They produce multiple images, strong distortions, and very
beautiful pictures of which a number can be found throughout this volume!

The two situations in which near-critical surface densities are
attained are in the cores of massive galaxy clusters, and in the cores
of galaxies (strictly speaking microlensing, described later, is also
a form of strong lensing). Both galaxy and cluster lensing are
discussed in detail in other contributions to this school.

Strong lenses are used in a number of applications: the main ones are
non-dynamical mass measurements of galaxies and of clusters,
determination of the Hubble constant, and studying distant galaxies with
the aid of the lensing magnification.

\subsection{Singular isothermal lens}

\begin{figure}
\plotone{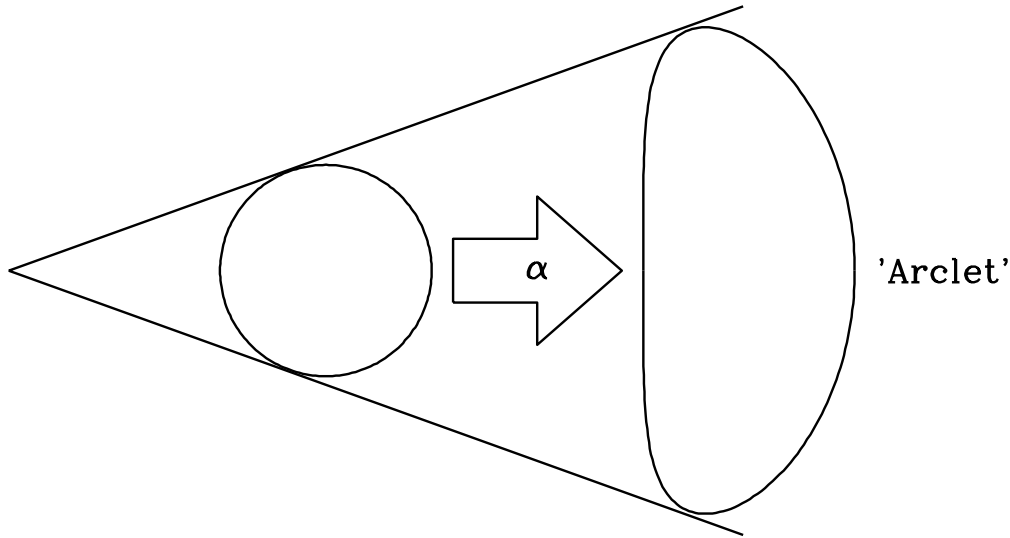}
\caption{Arclet formation by a singular isothermal sphere lens: a
circular source that is mapped outward by a constant deflection angle
is tangentially stretched into an arclet.}
\label{fig:arclet}
\end{figure}

A fiducial model for a galaxy or galaxy cluster lens is the
pseudo-isothermal sphere,
\begin{equation}
\psi(r)=\sigma^2\ln\left(a^2+r^2\right).
\end{equation}
It produces an asymptotically constant bending angle of 
\begin{equation}
\alpha={4\pi \sigma^2\over c^2} {s\over\left(a^2+s^2\right)}^{1/2}
\to 25\arcsec\left(\sigma\over1000\rm km\,s^{-1}\right)^2
\quad\hbox{at large $s$}.
\end{equation}
In this lens model, $\kappa$ and $(\gamma_1^2+\gamma_2^2)^{1/2}$ are
everywhere equal, and fall off as $s^{-1}$; on the critical line both
are equal to $\frac12$. The fact that the bending angle is constant
outside the core means that all images are displaced radially outward
by the same amount, and stretched tangentially along circles centered
on the lens. The resulting {\em arclets} surrounding the lens are a
generic feature of concentrated lenses, a manifestation of the lens's
outward `squeezing' of the sky (Fig.~\ref{fig:arclet}). They are
distinct from the {\em giant arcs} seen in elliptical lenses such as
the one in Fig.~\ref{fig:lensmap}: those are merging tangentially
distorted image pairs.

\subsection{The Hubble constant from time delays}

If one `solves' a lens system, this means that all angles between
observer, lens and source have been defined. However there is no
absolute distance scale that enters such a solution. The lens mass and
the distances between lens, source and observer can be rescaled in
proportion without any observable consequence to the images. However
such a rescaling does scale the Fermat potential, and hence the time
delay between images of the same source.

If a multiply-imaged source is variable, this offers the possibility
to measure this time delay directly, and hence to set such a scale.

In practice this method suffers from some difficulties. The lens needs
to be well-understood, which implies that either there is an
independent measurement of its mass profile (including the environment
of the lens itself)---the mass sheet degeneracy manifests itself
here. The time delay needs to be unambiguously established, requiring
long time series measurements, and feasible to monitor: this rules out
cluster lenses where the delay runs into centuries. Also, the relative
brightnesses of the images can be used as a further constraint on the
model; however the magnification depends on higher derivatives of the
potential and is therefore more sensitive to details. Furthermore, as
discussed in \S5 below, microlensing
may affect the images in different ways which are unrelated to the
large-scale lensing potential that determines the time delay.

The most robust systems for this work are four-image galaxy
lenses. They provide a useful number of constraints on the lens, so
that the potential may be sensibly constrained (although even here one
quickly runs out of parameters!). They also yield a number of
different time delay measurements, further tightening the model.

\begin{figure}
\plotone{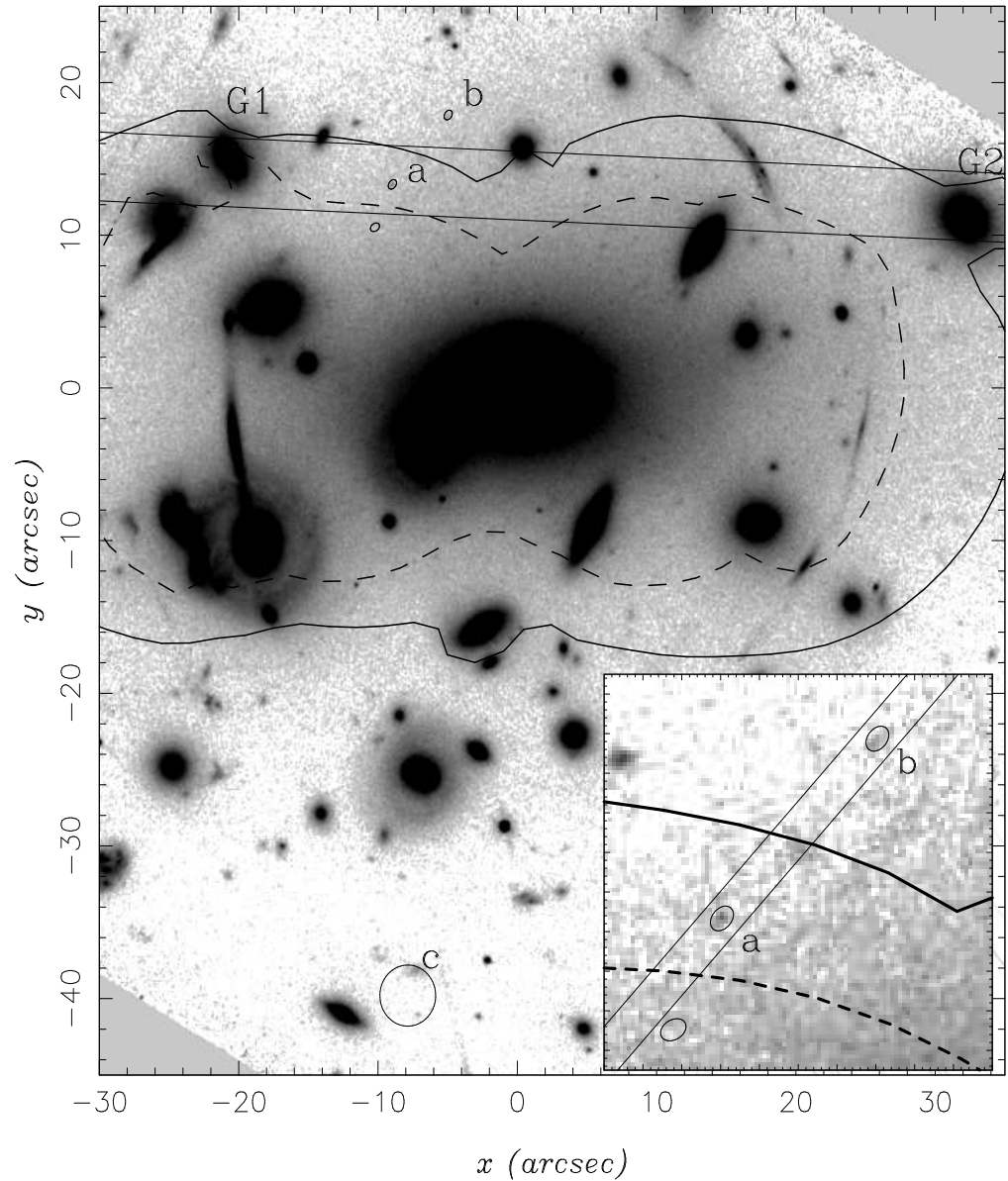}
\caption{The baby galaxy discovered behind Abell 2218. The source (a)
was discovered in a systematic survey of the critical curve region
(solid and dashed lines are the critical curves for source redshifts
5.5 and 1.1, respectively) in this cluster, and is a $z=5.5$
Ly-$\alpha$ emitter. Source (b) coincides with the predicted
counterimage of this source for a redshift of 5.5. Both images have
identical spectra. Had the source been at redshift 1.1, a second image
would have been visible below the corresponding critical line (shown
with the unlabelled ellipse).}
\label{fig:baby}
\end{figure}

\subsection{Clusters as Gravitational Telescopes}
Though their imaging properties are far from ideal, cluster lenses can
provide a useful peek at distant galaxies, magnifying detail that is
otherwise inaccessible. They also can boost faint sources into the
realm of detectability.

A nice example is provided by the 'baby galaxy' discovered behind
Abell 2218 by Ellis et al.~(2001). The cluster is one of the most
well-understood ones, with a number of multiple image systems for
which redshifts have been determined. By surveying the critical lines
of this cluster, a $z=5.56$ source was discovered which is magnified a
factor of 33 by the lens (Fig.~\ref{fig:baby}). Even so, the source is
barely visible on deep HST exposures! The spectrum shows a single
emission line at 800nm, but no detectable continuum. Using the lens
model, it was possible to predict where other images of the source
should be visible as a function of the redshift of the source. When
the line is identified with Ly-$\alpha$, such a second image was
indeed seen, and it proved to have an identical spectrum. The
gravitational telescope thus brought this tiny galaxy into view, and
the lensing was also instrumental in securing the redshift of this
source.

\section{Weak Lensing}

In weak lensing $\kappa, \gamma\ll1$. There is no multiple image
formation, only the slight distortion is under consideration in this
regime. Weak lensing is nevertheless a very powerful tool, which is a
result of two key facts: (i) it is possible to measure very small
distortions; and (ii) there is a nice inversion possible between the
observed distortion field and the lens mass distribution.

The details of these techniques are described in the review by Peter
Schneider elsewhere in this volume; here we sketch the principles.

\subsection{Measuring distortions}
\begin{figure}
\plotone{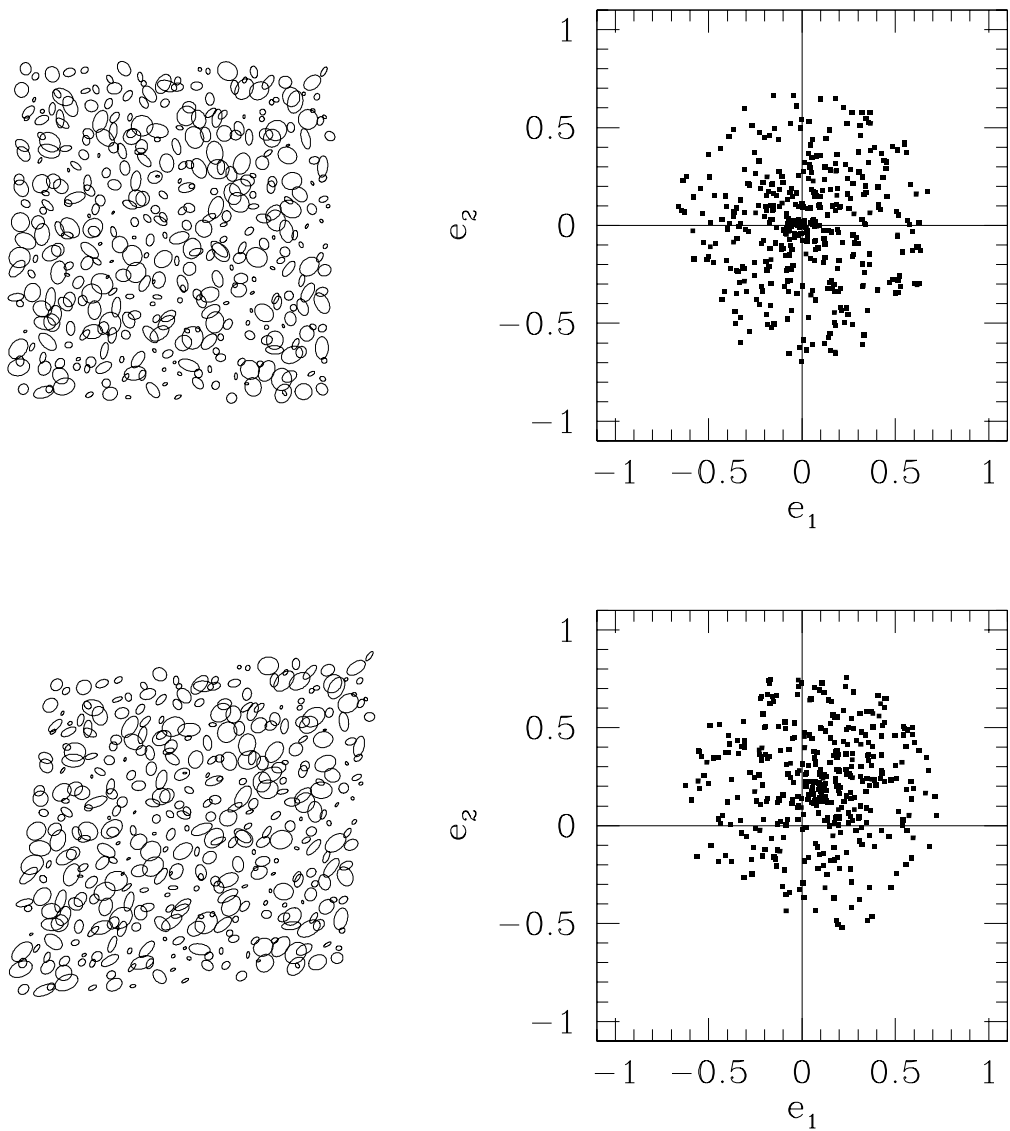}
\caption{An illustration of recovering the shear from a field of many
  randomy-oriented, elliptical galaxy images. On the left the galaxy
  images before (top) and after (bottom) applying a gravitational
  shear; on the right the corresponding distributions of image
  polarizations. In the unlensed case the polarizations cluster around
  zero, but a shear systematically moves the distribution.}
\label{fig:wkl}
\end{figure}

If we were in the fortunate position that all galaxies were
intrinsically round, then it would be simple to deduce the distortion
matrix of the lens mapping: the circular images would be sheared into
ellipses, and the size and direction of this shear would be easy to
read off from the shape and orientation of the galaxies. This ideal
can be approached by noting that while galaxies themselves come in
many, basically elliptical shapes, they are randomly oriented. The
{\em average} galaxy is thus round, in the sense that a stacking of a
large number of galaxy images would give a round blob. If there is
some distortion, this would affect all galaxy images equally, hence
also the summed image. Consequently we should be able to measure the
distortion from the way the blob is sheared.

In practice this is not the technique used to extract estimates of the
lensing distortion, but it illustrates that the information is
there. A different illustration of the same point is given in
Fig.~\ref{fig:wkl}, where an ensemble of elliptical images is sheared,
and the resulting systematic distortion uncovered by plotting the
distribution of image {\em polarizations}
\begin{equation}
\left(\!\!\begin{array}{c}e_1\\e_2\end{array}\!\!\right)
={1\over I_{11}+I_{22}}
\left(\!\!\begin{array}{c}I_{11}-I_{22}\\2I_{12}\end{array}\!\!\right)
\end{equation} 
where the $I_{ij}$ terms are different second moments $\int
f(x_1,x_2)x_ix_j\,dx_1dx_2$ of each galaxy's
light distributions $f(x_1,x_2)$ in the image plane. 

A delicate point in weak lensing is correction for all kinds of
atmospheric and instrumental effects, which often produce considerably
stronger distortions than the gravitational lensing. With existing
techniques it is possible to derived reliable distortions with an
accuracy of better than a few tenths of a percent. By obtaining
distortion estimates in various bins on an image, a distortion map can
be generated, which lays bare the effect of weak lensing by, for
example, a cluster of galaxies. Also statistical studies in the field
are now underway, in order to estimate the power spectrum of the
gravitational lensing distortions, which can be related to the power
spectrum of mass density fluctuations at intermediate redshifts.

\subsection{Mass reconstruction}
Once the distortion map has been made, the next step is to turn this
into information about the gravitational lens. This is possile,
because in the weak lensing regime, the distortion is a rather direct
masure of the gravitational shear $\gamma_i$ of the lens (though one
still needs to deal with the mass-sheet degeneracy, e.g. by setting
the average mass density at a large distance to zero). Using
eq.~\ref{eq:kapgam}, it is simple to show that the shear components
are related to $\kappa$, and hence the surface mass density
distribution in the lens, via
\begin{equation}
\gamma_{1,1}+\gamma_{2,2} = \kappa_1;\qquad 
\gamma_{2,1}-\gamma_{1,2} = \kappa_2.
\label{eq:massrecon}
\end{equation}
These equations, which are in fact redundant, allow the measured shear
field to be transformed into a $\kappa$ field, i.e., a picture of the
lensing mass distribution. Different techniques exist to perform this
inversion in practical situations, employing various kinds of
regularization, and handling boundary conditions in different ways,
but the principle is the same.

\section{Microlensing}
\label{sec:microlensing}
The final context in which gravitational lensing is encountered in
modern astronomy is {\em microlensing}. Microlensing is done by
compact lenses such as stars, and described by the point-lens
formalism. An important difference with weak and strong lensing is
that the multiple images produced by a microlens are usually not
resolved, but only the combined (enhanced) flux of both images is
seen. Because of this lack of resolution, microlensing can only be
detected when the lens-observer-source alignment changes, since this
alters the combined fluxes of the images. Detecting microlensing thus
involves monitoring sources for brightness variations.
\begin{figure}
\plotone{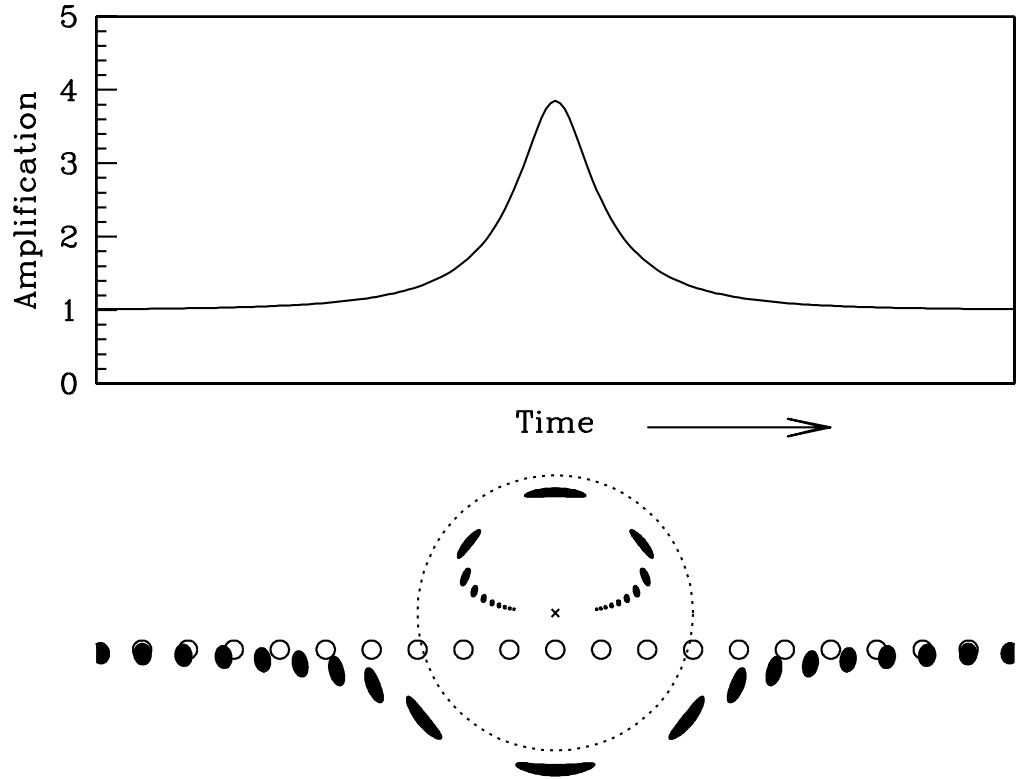}
\caption{The image configuration for point-mass lensing. The lens
  location is marked by the cross---this point is also the caustic for this
  lens. The corresponding critical curve (the Einstein circle) is
  shown as a dashed curve. Always two
  images are generated (the singularity breaks the
  odd-number-of-images rule), on either side of the lens. As the
  source tracks across the line of sight to the lens (open circles),
  the combined flux in both images (shown filled) rises and falls in a
  characteristic manner, as plotted in the top panel.}
\end{figure}

In quasars lensed by galaxies microlensing can occur as a result of
the grainy gravitational potential of the lens galaxy. This is known
as the high optical depth regime: microlensing in this context is not
well described by lensing due to a single point mass.

In a more local setting, on the other hand, microlensing is very rare
(for example, of order one star per million in the Galactic bulge or
in the Magellanic Clouds is microlensed at any time), and studying the
rates and characteristics of such microlensing events provides
important information on the number and masses of compact objects
along the line of sight.

The first surveys for microlensing were started in 1993, and the
results as well as future prospects are reviewed in detail in Wyn
Evans' article in this volume.

\section{Conclusions}

Gravitational lensing has matured into a standard tool of
astrophysics. It remains an attractive field because of the rather
clean concepts employed, the very different insight it offers into
mass determinations and geometrical measurements of the universe, the
occasional peek it grants into the distant universe, and the often
stunningly beautiful images one is forced to work with. As is well
illustrated in the rest of this volume, there is plenty more life in
lensing yet!

\label{page:last}

\end{document}